\documentclass[ amsmath, amssymb, aps, prb, twocolumn, superscriptaddress, floatfix ]{revtex4-1}%
\usepackage{graphicx} 
\usepackage{siunitx}
\usepackage{dcolumn} 
\usepackage{bm} 
\usepackage{xspace} 
\usepackage{placeins} 
\usepackage{tikz} 
\usepackage{amsmath}
\usepackage{natbib}
\usepackage{amsfonts}
\usepackage{amssymb}
\usepackage{upgreek}
\usepackage{textcomp}

\providecommand{\U}[1]{\protect\rule{.1in}{.1in}}

\setcitestyle{super}

\begin{document}
\title{Physical and one-dimensional properties of single crystalline La$_{5}$AgPb$_{3}$}
\author{Jannis Maiwald}
\affiliation{Quantum Matter Institute, University of British Columbia, Vancouver, BC, Canada}
\affiliation{Department of Physics and Astronomy, University of British Columbia, Vancouver, BC, Canada}
\email{jannis.maiwald@ubc.ca}
\author{M. C. Aronson}
\affiliation{Quantum Matter Institute, University of British Columbia, Vancouver, BC, Canada}
\affiliation{Department of Physics and Astronomy, University of British Columbia, Vancouver, BC, Canada}

\date{\today}

\begin{abstract}
We report here the properties of single crystals of La$_{5}$AgPb$_{3}$, which is a member of the $R_5MX_3$ ($R$ = rare earth, $M$ = transition metal or main group element, $X$ = Pb, Sn, Sb, In, Bi) family of chain-like compounds. Measurements of the electrical resistivity, specific heat and magnetic susceptibility are compared to the results of density functional calculations, finding that La$_{5}$AgPb$_{3}$ is a non-magnetic metal with moderate correlations. The analysis of the electrical resistivity and specific heat measurements highlight the importance of lattice vibrations in the material, while the calculated electron density suggests the presence of localized La and well-hybridized Ag atoms that extend along the c-axis in the La$_{5}$AgPb$_{3}$ structure. The temperature dependence of the magnetic susceptibility is consistent with a possible one-dimensional character of La$_{5}$AgPb$_{3}$, where the strength of correlations is much weaker than in one-dimensional conductors that have been previously reported.


\end{abstract}

\pacs{Valid PACS appear here}
\maketitle

\sisetup{range-phrase=--}

\section{Introduction}

Rare earth-based intermetallic compounds feature significant electronic correlations, which lead to a host of fascinating ordered or disordered phases, such as complex, at times frustrated, magnetism, superconductivity and quantum phase transitions\cite{Steglich2001,Johnston2010,Coleman2005}. Of particular interest are low-dimensional or geometrically frustrated materials, where quantum fluctuations can overwhelm the tendency to adopt ordered ground states. In these cases, the normal metallic behavior found in conventional conductors is replaced by ground states with unconventional excitations, such as fractionalized excitations. One-dimensional systems are of particular interest, due to the body of theoretical results that can be tested in real materials using inelastic neutron scattering \cite{lake2010confinement,mourigal2013fractional}, and angle resolved photoemission \cite{claessen2002spectroscopic,dudy2012photoemission}.

While most experimental work has been carried out so far on insulating spin chain systems, there is a pressing need to identify new materials where the physics of one dimension can be pursued in systems where electronic correlations are not strong enough to lead to localized electrons and insulating behavior. In three-dimensional materials, reducing the strength of correlations leads to a Mott transition or crossover where the localized electrons become itinerant, and the moments collapse with the onset of a metallic phase. Investigations into whether a corresponding scenario occurs in one-dimensional systems have been pursued largely in organic conductors \cite{kanoda2011mott}, and there is great interest in finding new classes of compounds where the interplay of low dimensionality and intermediate correlation strengths may lead to novel excitations and new quantum phases.

In seeking new correlated systems with quasi-one dimensionality, it is desirable to find a family of isostructural compounds where materials trends can be related to functionality, and where control over differing degrees of correlations is possible. The ternary compounds $R_5MX_3$ ($R$ = rare earth, $M$ = transition metal or main group element, $X$ = Pb, Sn, Sb, In, Bi) present a promising example of such a family, having the main structural motif of quasi-infinite linear chains composed of confacial triangular antiprisms \cite{Guloy1994}. These compounds form with a rich variety of magnetic and nonmagnetic atoms such as $R$ = Ce,Pr,Tb,Dy,Ho,Tm and $X$ = Mn,Fe,Co,Ni. Accordingly, existing experimental work has found exceptionally rich magnetic behavior, with evidence for both highly localized magnetic moments, and as well increasing hybridization and resulting Kondo compensation of those moments \cite{Rieger1968,Guloy1994,Gulay2005}. 

We focus here on La$_{5}$AgPb$_{3}$, which is a nonmagnetic member of this class of materials. Previously unstudied, we assess its potential as a benchmark system of one-dimensional chains of weakly correlated electrons, whose spin and angular momentum provide the only entities that might support magnetism. Electronic structure calculations support the proposal that La$_{5}$AgPb$_{3}$ may be one-dimensional, and as well suggest a variety of compositional variants where both transition metal and rare earth magnetism can be introduced. La$_{5}$AgPb$_{3}$ demonstrates Fermi liquid behavior in its low-temperature electrical resistivity and specific heat, with modest enhancements of the densities of states relative to values found in density functional theory (DFT) calculations. The temperature dependence of the magnetic susceptibility has a broad maximum, similar to those found in quasi-one dimensional metals.

\section{Methods}

Single crystals of La$_{5}$AgPb$_{3}$ were synthesized using a self-flux method. Elemental La (99.9\,\%), Ag (99.999\,\%) and Pb (99.999\,\%) in a ratio of 6:6:1 were placed in an alumina crucible set equipped with a strainer and a catch crucible, which was then sealed under $\sim$ 300\,mbar argon atmosphere in a quartz tube.
This assembly was subsequently heated at a rate of $\sim$ \SI{330}{\degreeCelsius\per\hour} to \SI{1030}{\degreeCelsius} where it remained for 3\,h to ensure adequate mixing of the reactants. Single crystals were precipitated out of solution by lowering the temperature at a rate of \SI{3}{\degreeCelsius\per\hour} down to \SI{930}{\degreeCelsius}. At that temperature, the tubes were removed from the furnace and were quickly placed upside down in a centrifuge, where the liquid flux was separated from the crystals by spinning at \SI{2000}{rpm} for about \SI{30}{\sec}.

Using this procedure we were able to grow needle-like single
crystals of La$_{5}$AgPb$_{3}$ for the first time. Typical dimensions
of the crystals are 4\,$\times$\,0.5\,$\times$\,0.5\,mm$^{3}$. The basal plane of the crystals is rectangular in shape with beveled edges. An image of the single crystal used for specific heat measurements is shown in the inset of Fig.~\ref{fig_xrd}.

We found La$_{5}$AgPb$_{3}$ to be air sensitive. The initially metallic luster of our single crystals turned to dull grey when exposed to air for a few minutes. Once the surface transformed there was no further change to the crystals, even after several days. However, an additional X-ray powder diffraction pattern taken on a powdered sample that had been exposed to air  for about two weeks showed that La$_{5}$AgPb$_{3}$ fully breaks down, as the pattern can no longer be indexed with the reported structure. The new pattern (not shown) indicates that that the main products of this breakdown are Lanthanum hydroxide (La(OH)$_3$), Lanthanum Lead (LaPb$_3$), Lead(II) oxide (PbO) and Silver oxynitrate (Ag$_7$NO$_{11}$). As a consequence the individual crystals were prepared for measurements under protective argon or helium atmosphere. Samples were only very briefly ($<$\SI{30}{\second}) exposed to air as they were loaded into the respective measurement instrumentation.

Powder x-ray diffraction (XRD) patterns were recorded with a Bruker D8 Advance diffractometer in the Bragg-Bretano configuration using a Cu cathode. The powder diffraction patterns were refined with the FullProf software suite. Energy-dispersive x-ray spectroscopy (EDX) was performed on a Philips XL-30 scanning electron microscope equipped with at Bruker xFlash 10\,mm$^2$ silicon drift detector. Measurements of the electrical resistivity and specific heat were carried out using a Physical Property Measurements System (PPMS) from Quantum Design equipped with a He$^{3}$/He$^{4}$ dilution refrigerator insert. Measurements of the magnetic susceptibility were performed using a Magnetic Property Measurements System 3 (MPMS3) also from Quantum Design.

Band structure calculations were performed using the Linear Augmented Plane Wave code WIEN2k\cite{Wien2k} (Version 21.1) on the reported crystal structure with space group $P6_3/mcm$. The gradient-corrected density functional of Ref.~\onlinecite{Perdew1996} (PBE) was used in all calculations. The self-consistent field cycles were computed using an optimized basis set size and k-mesh of $R_\text{MT}k_\text{max} = 9$ and 5000 $k$ points, respectively.

\section{Structural \& Physical Properties} \label{section_structure}
\subsection{Sample characterization}

\begin{figure}[ptb]
\centering
\includegraphics[width=0.95\linewidth]{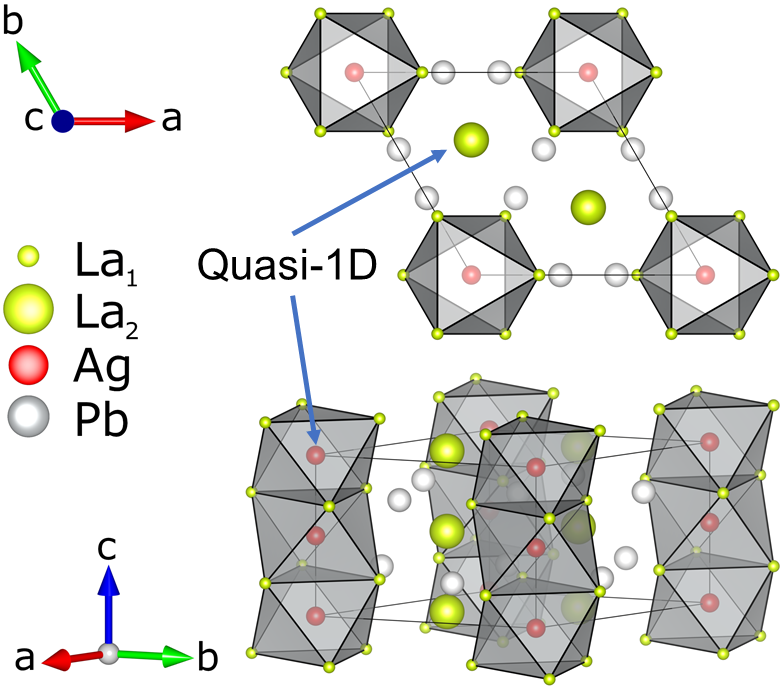}
\par
\caption{ The refined crystal structure of La$_{5}$AgPb$_{3}$. The unit cell (thin grey lines) is shown in (top) top view and (bottom) perspective view. La atoms on the La$_1$ (6g) site were reduced in size for clarity. Arrows indicate quasi-one-dimensional arrangements of La$_2$ and Ag, respectively.}
\label{fig_structure}
\end{figure}

The ternary compounds $R_5MX_3$ ($R$ = rare earth, $M$ = transition metal or main group element, $X$ = Pb, Sn, Sb, In, Bi) form in a hexagonal crystal structure with space group $P6_3/mcm$ (193) (Fig.~\ref{fig_structure})\cite{Tran2007,Goruganti2008,Goruganti2009}. These systems feature quasi-one-dimensional arrangements of their constituent $M$ atoms on the 2b sites and $R$ atoms on the 4d sites that extend along the crystallographic c axis. The $R$ atoms on the 6g site shield the $M$ atoms from the $X$ atoms by forming irregular octahedra (confacial trigonal antiprisms) around the $X$ atoms. These materials form in the hexagonal Hf$_5$CuSn$_3$ structure, more broadly known as the Ga$_4$Ti$_5$ structure type, which is distinguished from the more populous Mn$_5$Si$_3$ structure type by the occupancy of the $M$ site \cite{Rieger1968,Guloy1994,Gulay2005}.

La$_{5}$AgPb$_{3}$ forms in the Hf$_5$CuSn$_3$ structure type, as depicted in Fig.~\ref{fig_structure}. Layers of La (4d-site, La$_2$) and Pb are separated by layers of La (6g-site, La$_1$) and Ag that are stacked along the crystallographic c axis. The Hf$_5$CuSn$_3$ structure type is differentiated from the Mn$_5$Si$_3$ structure by the presence of the Ag atoms occupying the corners of the unit cell, vacant in the Mn$_5$Si$_3$ structure. 

The Ag atoms on the 2b sites, as well as the La atoms on the 4d sites form linear chains. The La atoms on the La$_1$ site form irregular octahedra around the Ag atoms on the 2b sites, effectively shielding them from the Pb atoms. This arrangement suggests that Ag-Pb bonding is weak, as expected considering the limited solid solubility of Ag and Pb. This type of behavior can frequently be observed when introducing a third element to an 'antagonistic pair' of elements\cite{canfield2019new}. More importantly this structure in La$_{5}$AgPb$_{3}$ contributes to the formation of quasi-one-dimensional arrangements of the Ag and La$_2$ sites along the crystallographic c axis (Fig.~\ref{fig_structure}). Replacing the respective elements with magnetic elements in the future could provide a great opportunity to determine whether one-dimensional magnetism can be induced in this class of materials. 

\begin{table}[b]
\caption{Results -- EDX analysis \& Rietveld refinement.}
\label{tab_refinement}
\begin{ruledtabular}
		\begin{tabular}{llll}
		Composition [at.\si{\percent}] & La & Ag & Pb \\
		\hline
		& & &\\
		EDX Average & 57.8(3) & 10.0(1) & 32.2(4)\\
		Nominal & 55.6& 11.1 & 33.3		 \\
		& & &\\
		Lattice Parameters & a [\si{\angstrom}] & c [\si{\angstrom}] & V [\si{\angstrom\tothe{3}}] \\
		\hline
		& & &\\
		La$_{5}$AgPb$_{3}$ & 9.7730(7) & 6.8900(7) & 569.91 \\
		Ref.~\onlinecite{Guloy1994} & 9.560(1) & 7.037(2) & 556.97 \vspace{2.5mm} \\
		Atomic Coordinates & x & y & z \vspace{1.5mm} \\
		La$_1$ (6g)	 & 0.2710(5) & 0.00000 & 0.25000\\
		La$_2$ (4d)	 & 0.33330 & 0.66670 & 0.00000\\
		Ag \ (2b) & 0.00000 & 0.00000 & 0.00000 \\
		Pb \ (6g) & 0.6197(4) & 0.00000 & 0.25000 \\

		\end{tabular}
	\end{ruledtabular}
\end{table}

\begin{figure}[ptb]
\centering
\includegraphics[width=0.95\linewidth]{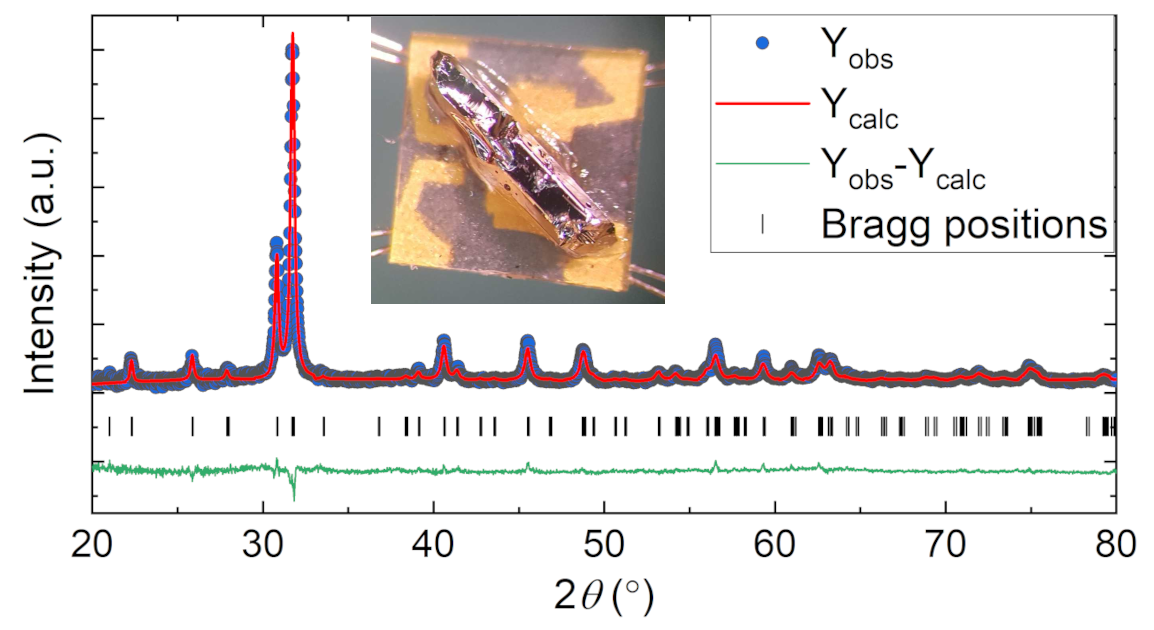}
\par
\caption{Measured powder XRD pattern of as grown La$_{5}$AgPb$_{3}$, and its refinement using the reported hexagonal structure with space group $P6_3/mcm$. Bragg peak positions are indicated. (insert) Optical microscope image of a La$_{5}$AgPb$_{3}$ single crystals (length: $\sim$4\,mm) mounted on a heat capacity platform.}
\label{fig_xrd}
\end{figure}

The composition of as-grown single crystals of La$_{5}$AgPb$_{3}$ was confirmed using Energy Dispersive X-Ray analysis (EDX). The mean composition, averaged over 19 measurement points, is given in Table~\ref{tab_refinement}. The data all agree with the nominal composition of La$_{5}$AgPb$_{3}$, given a typical EDX resolution of 1-2\,\si{\percent}. 

The crystal structure of the synthesized La$_{5}$AgPb$_{3}$ crystals was verified by means of x-ray powder diffraction.
The diffraction pattern of powdered single crystals of La$_{5}$AgPb$_{3}$ and its Rietveld refinement are compared in Fig.~\ref{fig_xrd}. All recorded diffraction peaks can be indexed within the hexagonal structure reported for polycrystals of La$_{5}$AgPb$_{3}$\cite{Rieger1968,Guloy1994}. The absence of extrinsic diffraction peaks rules out the presence of crystalline impurity phases with concentrations larger than $\simeq$\SI{1}{\percent}. The refined lattice parameters and atomic coordinates are presented in Table~\ref{tab_refinement}. The former show about a \SI{\pm 2}{\percent} deviation from reported literature values. We attribute this difference to differences between polycrystalline samples and our single crystals. Our XRD measurements indicate that we have successfully grown stoichiometric high-quality single crystals of La$_{5}$AgPb$_{3}$.

\subsection{Density functional calculations of the electronic structure} \label{DFT}

\begin{figure*}[ptb]
\centering
\includegraphics[width=\linewidth]{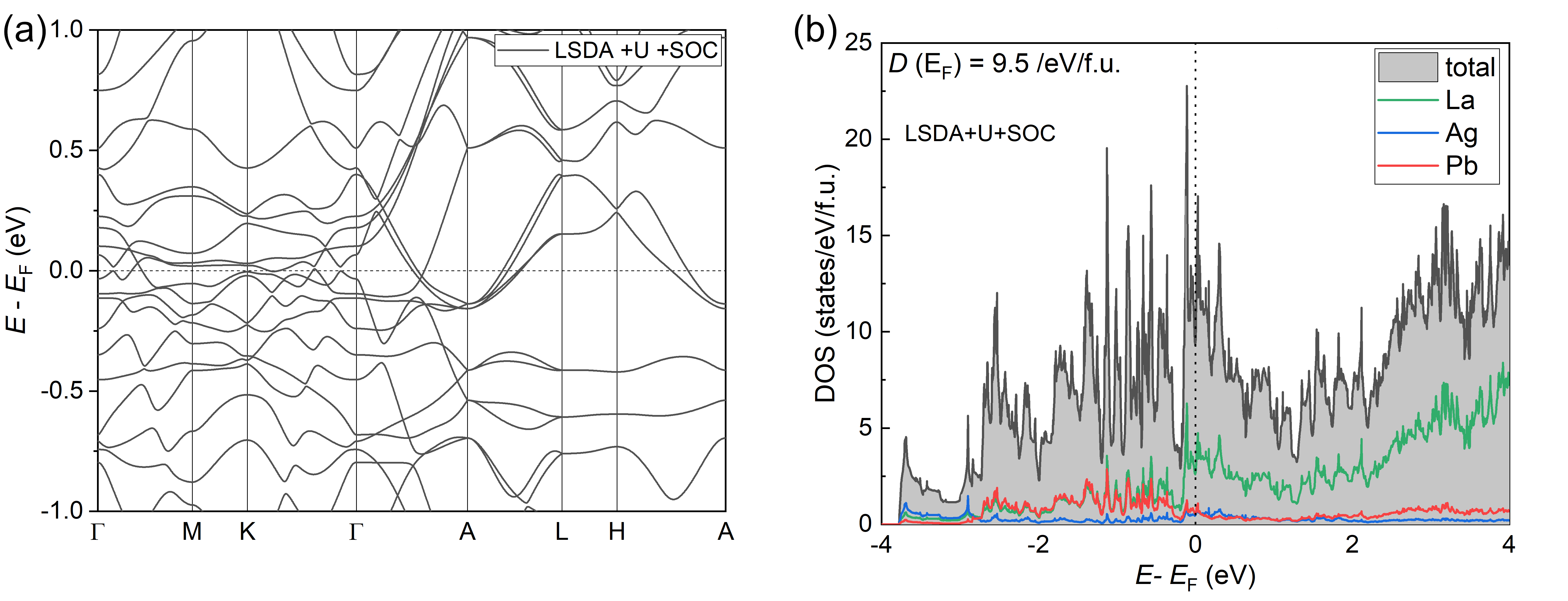}
\caption{Density functional calculation results for La$_{5}$AgPb$_{3}$. (a) Electronic band structure plotted along some high-symmetry directions in the hexagonal unit cell. (b) Electronic density of states as a function of energy $E$ relative to the Fermi energy $E_\text{F}$, with $\mathcal{D}(E_\text{F})=$ 9.5 states/eV/f.u.}%
\label{fig_fermi}%
\end{figure*}

\begin{figure}[ptb]
\centering
\includegraphics[width=\linewidth]{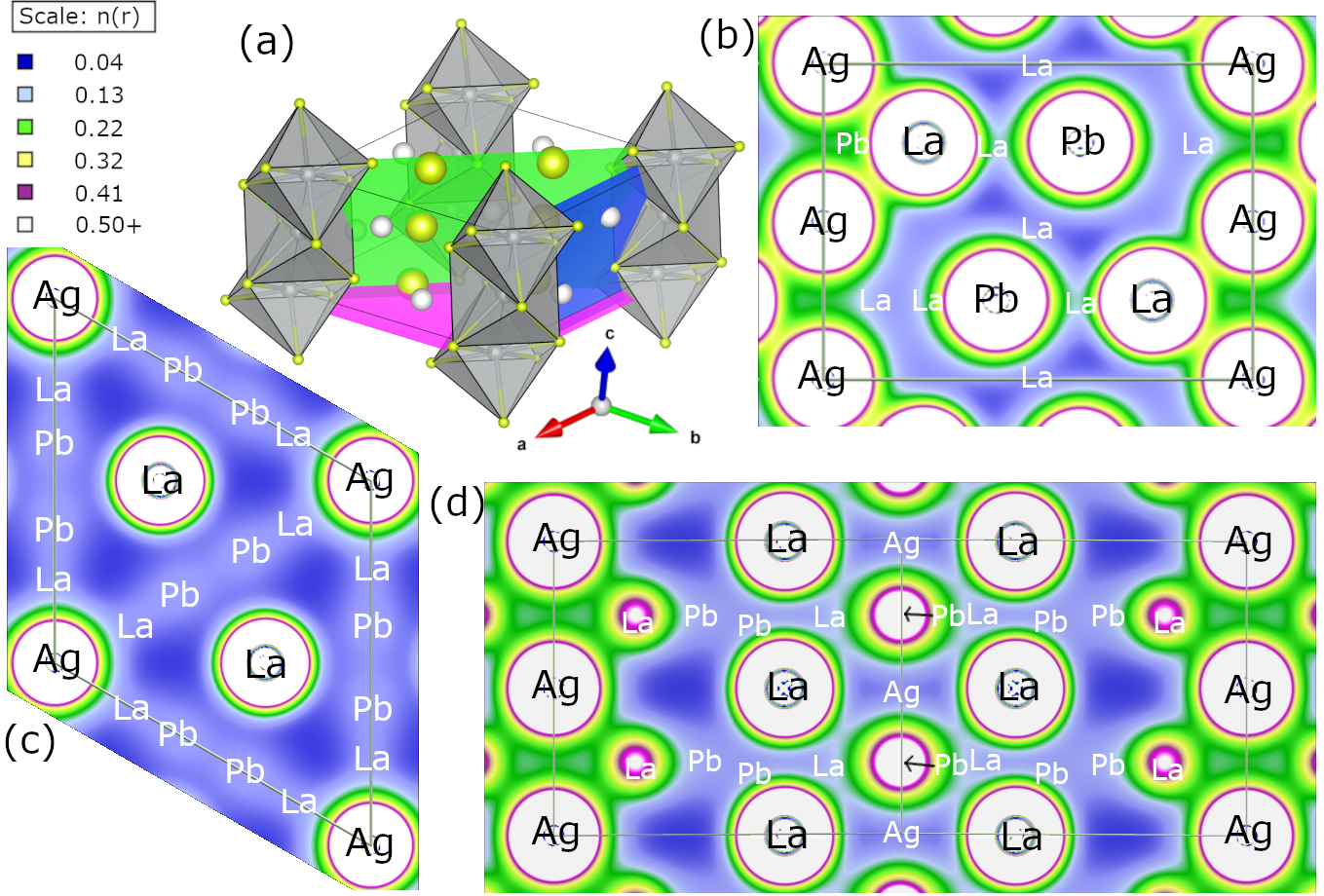}
\caption{Density of valence electrons in La$_{5}$AgPb$_{3}$. Panels (b-d) show cuts through the (010), (110) and (001) plane of the hexagonal unit cell (thin grey lines), respectively. (a) The corresponding lattice planes are highlighted in the unit cell in blue, green and red, respectively. Note that atoms labels in white indicate atoms behind and in front of the depicted planes. The linear color map is truncated at 0.5 \si{electrons\per\angstrom\tothe{3}}.}
\label{fig_densities}%
\end{figure}

We have performed \emph{ab~initio} band structure calculations based on the local-spin-density approximation (LSDA) using the experimentally determined hexagonal crystal structure of La$_{5}$AgPb$_{3}$. We have included spin-orbit coupling (SOC) for all elements, as well as an on-site Coulomb interaction ($U$) for La, to minimize the influence of the (empty) La 4$f$ states on the density of states (DOS) at the Fermi energy E$_{F}$. The Coulomb and exchange parameters were set to $U=7.5$\,eV and $J=0$, respectively.
The calculated DOS and band structure of La$_{5}$AgPb$_{3}$ are depicted in Fig.~\ref{fig_fermi}. 

These calculations indicate that La$_{5}$AgPb$_{3}$ is a metal, with five bands crossing the Fermi level [Fig.~\ref{fig_fermi}(a)], leading to a robust density of states at the Fermi level $\mathcal{D}(E_\text{F})=$ 9.5 states/eV/f.u. [Fig.~\ref{fig_fermi}(b)]. Most of $\mathcal{D}(E_\text{F})$ originates from electrons in the interstitial space between the muffin tin spheres, while smaller and more localized contributions from La, Ag and Pb states make only a negligible contribution to the density of states.

We have also calculated the density of valence electrons $n(r)$ of La$_{5}$AgPb$_{3}$. Representative cuts of $n(r)$ through the (010), (110) and (001) planes are depicted in Fig.~\ref{fig_densities}, and a video of the three-dimensional density can be found in the Supplemental Material.

Figure~\ref{fig_densities}(a) shows different planes in the crystal structure of La$_{5}$AgPb$_{3}$. Starting in Fig.~\ref{fig_densities}(b), we see the projection of $n(r)$ in the (010) plane, which is on the side of the unit cell (blue plane). The most overlap occurs between Ag and La atoms, forming connected paths along the crystallographic c direction, while the Pb atoms in the center are electronically well isolated in this plane. Only minute hybridization between the La$_1$ and Pb atoms is visible. The quasi-one-dimensional bonding along the c axis between the Ag and La$_1$ atoms is even more apparent from the cut through the center of the unit cell in Fig.~\ref{fig_densities}(d) (green plane). Since the La$_1$ atoms do not lie in the green plane, their density is only visible as red blobs close to the Ag atoms on the right and left hand sides of the unit cell. The La$_2$ atoms in the center of the plot, however, are electronically isolated, as reflected in the plot in panel~(c) (pink plane), where only residual density from the out-of-plane atoms is visible (white bands) between the isolated in-plane La$_2$ and Ag atoms.

It is evident from these plots that the vast majority of the valence electron density is localized around the constituent atoms (see white-colored circles in Fig.~\ref{fig_densities}) with very little density in the interstitial regions. The strongest orbital overlap visible occurs between the irregular La$_1$ octahedra and the enclosed Ag atoms, while the remaining atoms seem electronically well separated from each other, particularly the La$_2$ chains along the c axis. 

The results of the valence electron density are suggestive of quasi-one-dimensional behavior in La$_{5}$AgPb$_{3}$. They imply that substitution of Ag with a magnetic element would likely be the most promising route towards one-dimensional magnetism in this class of materials, since the substitution should be relatively straight forward. Replacing the more isolated La$_2$ atoms would also be interesting, but is much more challenging experimentally, given the presence of the second crystallographic La site in the material. Based on these findings, we hypothesize that magnetic substitution of Ag should lead to more delocalized, itinerant magnetism, while magnetic substitution of the La$_2$ site should lead to well isolated local moments. 

 Reports of electride behavior in compounds with similar composition and crystal structure\cite{Liu2020electrides,Zhu2019computational}, and in particular in the Mn$_5$Si$_3$ structure type \cite{Liu2020electrides}, have been published recently. However, we do not see any obvious aggregation of electrons in the interstitial space in La$_{5}$AgPb$_{3}$, which is the defining characteristic of electride materials. We suggest that this may be because the space in the corners of the unit cell, where such an aggregation is expected to occur (see e.g. Fig.~S2 in Ref.~\onlinecite{Zhu2019computational}), is occupied by Ag atoms in La$_{5}$AgPb$_{3}$, unlike the case of the Mn$_5$Si$_3$ where this site is unoccupied. In any case, our electronic structure calculations do not suggest that La$_{5}$AgPb$_{3}$ is an electride. 

\begin{figure}[ptb]
\centering
\includegraphics[width=\linewidth]{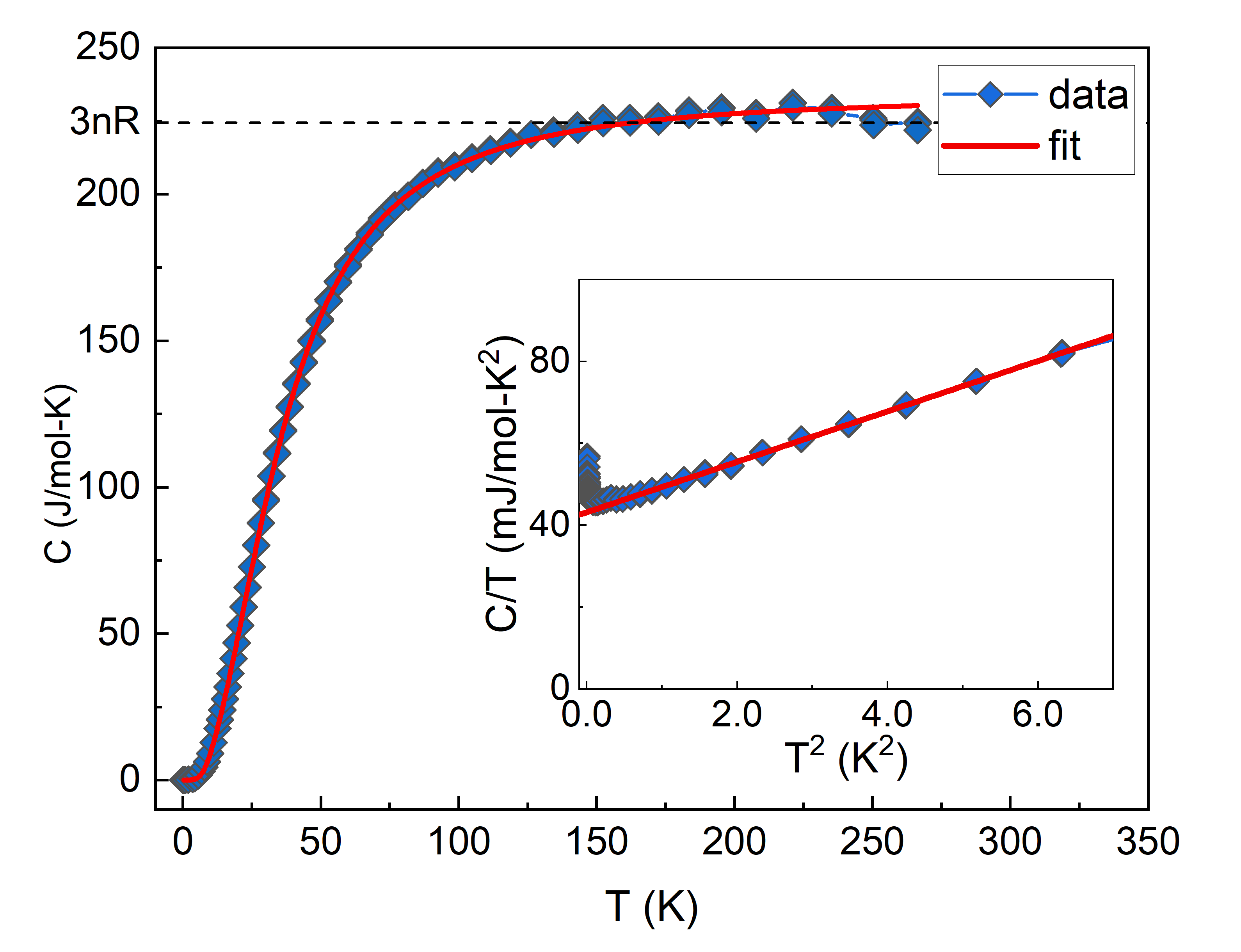}
\caption{Specific heat $C(T)$ of La$_{5}$AgPb$_{3}$ as a function of
temperature $T$ between \SIrange{0.05}{275}{\kelvin}. The red line is a fit to
the data using the Debye-Einstein model defined in the main text. 3$nR$ denotes the
Dulong-Petit constant with n=9 (dashed line). (Inset) Specific
heat $C/T$ as a function of $T^{2}$ between \SIrange{0.05}{2.5}{\kelvin}. The solid
red line is a fit to the expression: $C(T)=\gamma T+\beta T^{3}$.}%
\label{fig_hc}%
\end{figure}

\subsection{Specific Heat}

The specific heat $C(T)$ of La$_{5}$AgPb$_{3}$ was measured over the temperature range between \SIrange{0.05}{275}{\kelvin} (Fig.~\ref{fig_hc}), where $C(T)$ rises monotonically for temperatures as large as $\simeq$ 250 K. The Debye model gives a poor fit to the overall temperature dependence of $C(T)$, indicating that acoustic phonons alone are not sufficient to describe the specific heat of La$_{5}$AgPb$_{3}$. We have modelled the measured specific heat $C$ using a Debye-Einstein model that includes both acoustic and optical phonons:
\begin{equation} \label{eq_debye_einstein}
C(T)= m_1\cdot C_{\text{D}}\left(T/T_{\text{D}}\right)\ + m_2\cdot C_{\text{E}}(T/T_{\text{E}}).
\end{equation}
The weighting factors $m_i$ with $i=1,2$ enforce the requirement that the acoustic and optical modes must together satisfy the Dulong-Petit limit ($3nR$) at sufficiently large $T$. Hence, the sum of the weighting factors $\sum m_i$ should approximate the number of atoms per formula unit\cite{Woodfield1999}. We assume in our formula in Eq.~\eqref{eq_debye_einstein} that the acoustic modes are described by the Debye term, while the single Einstein mode provides an empirical equivalence of the spectrum of optical modes. The contributions to the specific heat are given by the Debye-Integral and an Einstein term
\begin{align*}
C_{\text{D}}\ & =9R\ \left( \frac{T}{T_{\text{D}}%
}\right)^{3}\int_{0}^{\frac{T_{\text{D}}}{T}}\frac{x^{4}}{(e^{x}%
-1)(1-e^{-x})}dx,\\
C_{\text{E}}\ & =3R\ \left( \frac{T_{\text{E}}%
}{T}\right)^{2}\frac{1}{(e^{T_{\mathrm{E}}{}/T}-1)(1-e^{-T_{\mathrm{E}}{}%
/T})},
\end{align*}
with universal gas constant $R$ and Debye and Einstein temperatures $T_{\text{D}}$ and $T_{\text{E}}$, respectively. The data are well described with the following parameters:
$m_1=\SI{7.9\pm0.1}{}$, $m_2=\SI{1.5\pm0.1}{}$, $T_{\text{D}}\ =\SI{159\pm1}{\kelvin}$ and $T_{\text{E}}\ =\SI{52\pm2}{\kelvin}$.

We extracted the Sommerfeld coefficient $\gamma$ and $T_{\text{D}}$ from a linear fit to the low-temperature $C/T$ for temperatures between \SIrange{0.7}{2.6}{\kelvin} (see inset of
Fig.~\ref{fig_hc}) using the expression: $C(T)=\gamma T+\beta T^{3}$, with $\beta=12\pi
^{4}/5\cdot nN_{\text{A}}k_{\text{B}}/T_{\text{D}}^{3}$. We get $\gamma
=\SI{43.08\pm0.07}{\milli\joule\per\mole\per\kelvin\tothe{2}}$ and $T_{\text{D}%
}\ =\SI{141.5\pm0.2}{\kelvin}$, with the latter being in reasonable agreement with the value found above. The Sommerfeld coefficient corresponds to $\mathcal{D}(E_\text{F}) = \SI{18.27\pm0.03}{states\per\electronvolt\per \text{f.u.}}$ at the Fermi level. This value is $\sim \SI{90}{\percent}$ bigger than the density of states determined by the DFT calculations $\mathcal{D}(E_\text{F}) = \SI{9.5}{states\per\electronvolt\per \text{f.u.}}$ This comparison indicates that electronic correlations are appreciable in La$_{5}$AgPb$_{3}$, leading to a moderate mass enhancement that is in line with observations in similar materials with the same crystal structure, such as La$_5$Ge$_3$ (\SI{33.6}{\milli\joule\per\mole\per\kelvin\tothe{2}}) and La$_5$Si$_3$ (\SI{27.0}{\milli\joule\per\mole\per\kelvin\tothe{2}})\cite{Gorbachuk1998}, but not La$_5$Pb$_3$ (\SI{6.5}{\milli\joule\per\mole\per\kelvin\tothe{2}})\cite{Demela}. Finally, we attribute the increase in $C$ that is visible only in a plot of $C/T(T^2)$ below \SI{0.75}{\kelvin} to a nuclear Schottky anomaly, likely associated with the large nuclear spin $I=\text{7/2}$ of $^{139}$La. We, therefore, limited the fit range in the above analysis to temperatures above \SI{0.75}{\kelvin}.

\subsection{Electrical Resistivity}

\begin{figure}[ptb]
\centering
\includegraphics[width=\linewidth]{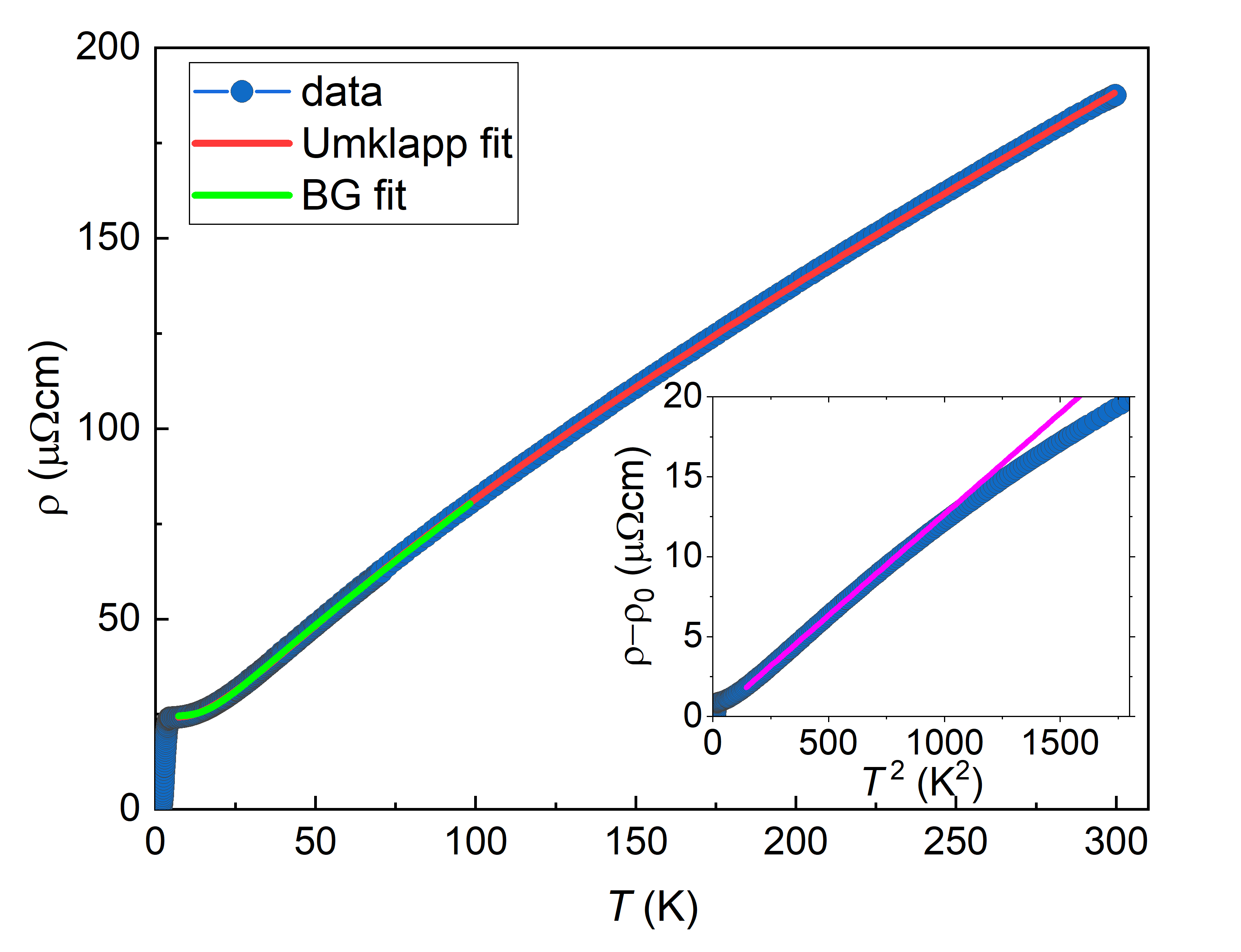}
\caption{Temperature dependence of the electrical resistivity $\rho(T)$ of La$_{5}$AgPb$_{3}$, with the measuring current $I$ applied along the crystallographic c axis. The data (blue symbols) are well described by the modified Bloch-Gr\"uneisen model with an added exponential term to account for phonon-assisted Umklapp processes as discussed in the main text (red line). The unaltered Bloch-Gr\"uneisen fit (green symbols) is also shown. (insert) $T^2$ temperature dependence of $\rho(T)$. A linear fit between \SIrange{12}{28}{\kelvin} is shown (magenta line).}
\label{fig_res}
\end{figure}

The electrical resistivity $\rho(T)$ of La$_{5}$AgPb$_{3}$ (Fig.~\ref{fig_res}) is of order \si{10\tothe{2}\mu\Omega\centi\meter} at room temperature, indicating metallic behavior in agreement with our DFT results. Accordingly, $\rho(T)$ decreases monotonically with decreasing temperature, initially being sub-linear but becomes linear between 100\,K and 35\,K before it flattens out with a residual value of $\SI{24}{\mu\Omega\centi\meter}$ at 5\,K. This results in a residual resistivity ratio of $\rho(295\,\text{K})/\rho(5\,\text{K}) = 8$, indicating sizable scattering at low temperatures. 
Below 4.2\,K the resistivity drops sharply, indicating a transition into a superconducting state. However, the broad superconducting transition is not complete, as the resistivity remains nonzero down to the lowest measured temperature of 1.8\,K. No evidence for a superconducting transition was found in the specific heat or magnetization measurements of our crystals. We believe that a thin layer of LaPb$_3$ with a superconducting transition temperature $T_\text{c}=4.18\,\text{K}$ \cite{Welsh1975} is responsible. Since LaPb$_3$ is a decomposition product of La$_{5}$AgPb$_{3}$, we speculate that it formed on the surface as the sample was briefly exposed to air as it was transferred into the measurement device. No indications for the presence of LaPb$_3$ have been observed in either our EDX or XRD analysis.

The resistivity in the temperature range \SIrange{5}{100}{\kelvin} is well described by the Bloch-Gr\"{u}neisen law:
\begin{equation}
\rho(T)= \rho_{0}+\rho_{\mathrm{D}}(T), \label{BG}
\end{equation}
\begin{equation}
\rho_{\mathrm{D}}
(T)=A\left(\frac{T}{T_{\mathrm{D}}}\right)^{5}\int_{0}^{T_{\mathrm{D}}/T}\frac{x^{5}dx}{\left(e^{x}
-1\right)\left(1-e^{-x}\right)},
\label{rD}
\end{equation}
where $A=\SI{254.6\pm0.6}{\mu\Omega\centi\meter}$, $T_{\mathrm{D}}=\SI{105.2\pm0.2}{\kelvin}$ and $\rho_0 = \SI{24.49\pm0.02}{\mu\Omega\centi\meter}$. This indicates that the scattering of weakly correlated quasiparticles from acoustic phonons is dominant in this temperature range. The Debye temperature obtained from this analysis is about \SI{25}{\percent} smaller than that deduced from our specific heat data. Adding a term $\rho_\text{el} (T) = AT^2$ to account for electron-electron scattering within Fermi liquid theory did not improve the fit, and in fact the $A$ parameter was refined to zero. However, plotting $\rho(T)$ vs $T^2$ reveals a small range of temperatures between \SIrange{12}{28}{\kelvin} where $\rho$(T) obeys a quadratic temperature dependence $\rho$(T)=$\rho_{0}$+AT$^{2}$ (Fig.~\ref{fig_res}, inset), reflecting the dominance of electron-electron scattering at the lowest temperatures. We obtain $A = \SI{1.265\pm0.002}{10\tothe{-2}\mu\Omega\centi\meter\per\kelvin\tothe{2}}$, leading to a Kadowaki–Woods ratio of $R_\text{KW} = A/\gamma^2 = \SI{6.8}{10\tothe{-8}\Omega\meter\cdot\mol\tothe{2}\kelvin\tothe{2}\per\joule\tothe{2}}$, which would put La$_{5}$AgPb$_{3}$ in the vicinity of metals like Sr$_2$RuO$_4$ and Rb$_3$C$_{60}$ with moderate correlations\cite{Jacko2009}.

In an attempt to describe $\rho$(T) over the full measured temperature range, we added an Einstein term, $\rho
_{E}(T)=(B/T)\ /(e^{T_{\mathrm{E}}{}/T}-1)(1-e^{-T_{\mathrm{E}%
}{}/T}),$ to the model\cite{Cooper1974}, as in the analysis of the specific heat. This did not improve the quality of the fit. However, adding an exponential term, $B\exp(-T_{0}/T),$ that accounts for Umklapp processes assisted by a specific phonon with an energy $T_{0}$~\cite{Woodard1964,Milewits1976}, gives an excellent fit to the data (Fig.~\ref{fig_res}, red line) over the entire measured temperature range with the following parameters: $A=\SI{151.6\pm0.8}{\mu\Omega\centi\meter}$, $T_{\mathrm{D}}=\SI{83.5\pm0.4}{\kelvin}$, $B=\SI{41.5\pm0.3}{\mu\Omega\centi\meter}$, $T_\text{0}=\SI{110.3\pm0.3}{\kelvin}$ and $\rho_0=\SI{24.14\pm0.02}{\mu\Omega\centi\meter}$. It is noteworthy that this procedure requires a Debye temperature that is smaller by a factor of 1/2, while $T_\text{0}$ is larger by a factor of 2 compared to the values for $T_{\mathrm{D}}$ and $T_{\mathrm{E}}$ extracted from our specific heat data, respectively. We have recently observed similar behavior in superconducting La$_2$Ni$_2$In \cite{Maiwald2020}.

\subsection{Magnetic Susceptibility}

\begin{figure}[ptb]
\centering
\includegraphics[width=\linewidth]{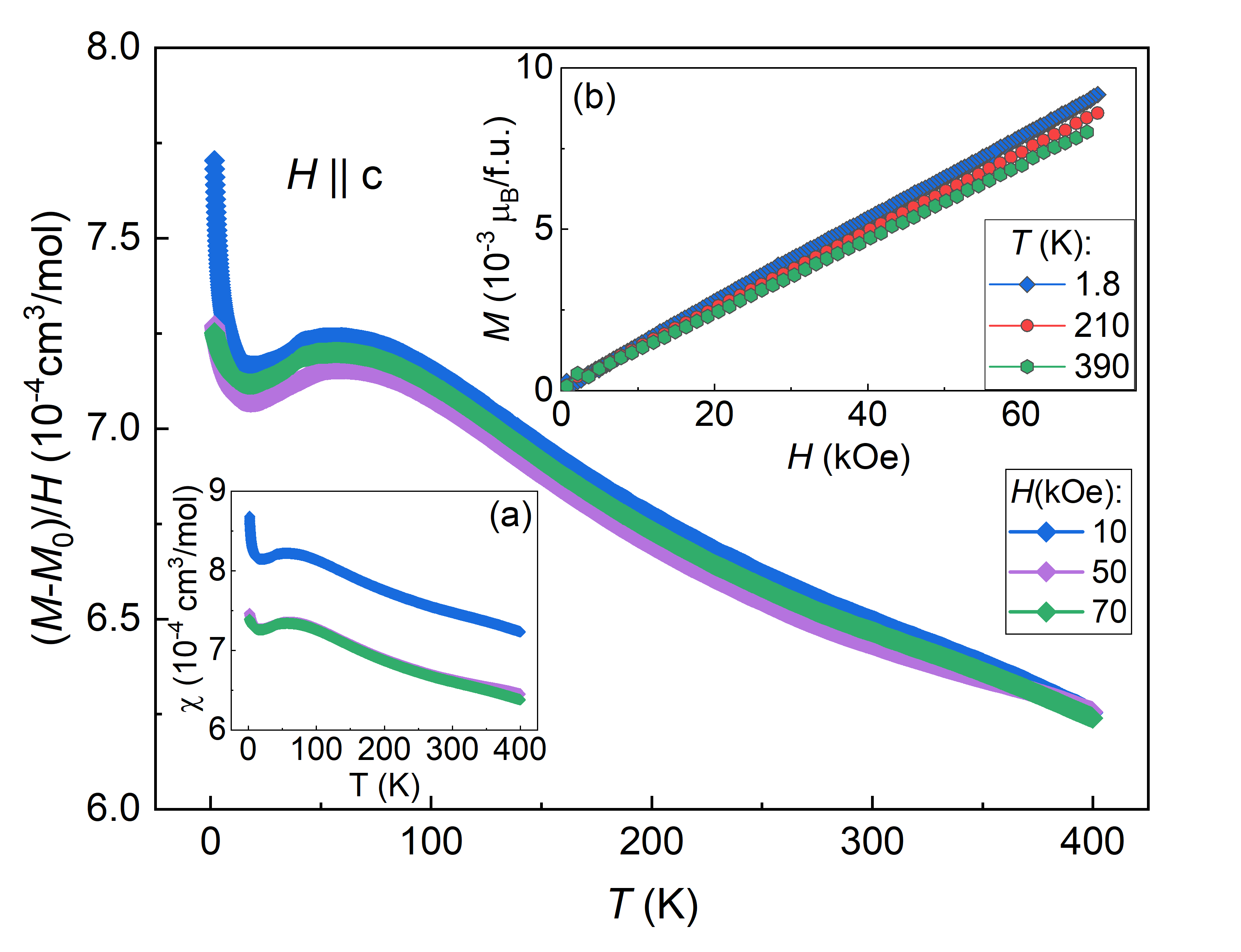}
\caption{
(a) The temperature dependencies of the magnetic susceptibility $\chi$(T), measured in fields of 10\,kOe (blue), 50\,kOe (violet) and 70\,kOe (green) that were applied parallel to the crystallographic c-axis. Main panel: Subtracting a constant magnetization M$_{0}$=0.98 emu/mol removes the field dependence of the dc magnetization. (b) The field dependencies of the magnetization $M$ measured at different temperatures.
}
\label{fig_susT}%
\end{figure}

The dc magnetic susceptibility $\chi(T)$ of La$_{5}$AgPb$_{3}$ with the magnetic field $H$ applied along the crystallographic c direction is only weakly temperature dependent over the measured temperature range \SIrange{1.8}{400}{\kelvin} [Fig.~\ref{fig_susT}(a)]. $\chi(T)$ increases slightly with decreasing temperature until $\sim$\SI{60}{\kelvin} where a broad maximum forms. A rapid increase indicative of a Curie-Weiss tail, presumably from paramagnetic impurities, is found below 18\,K. 

$\chi(T)$ exhibits a weak field dependence, particularly in smaller magnetic fields [Fig.~\ref{fig_susT}(a)]. This indicates the presence of a spontaneous magnetization $M_0$, likely introduced by a tiny amount of ferromagnetic impurities in the sample. Subtracting a small constant $M_0 = \SI{0.98}{emu/mol} \approx \SI{1.76}{\times10\tothe{-4}\mu_\text{B}/\text{f.u.}}$ from the data makes the measurements taken at different fields collapse on top of each other (Fig.~\ref{fig_susT}, main panel), corroborating this hypothesis. Taking $M_0 \sim \SI{1}{\mu_\text{B}/atom}$ we can estimate this ferromagnetic impurity concentration $n_\text{imp} = \SI{0.002}{\percent}$ present in the sample. This is far below the sensitivity limits of most analytical methods, including XRD and EDX. 

However, the spontaneous moment $M_0$ does not explain the Curie-tail visible in Fig.~\ref{fig_susT} at the lowest temperatures. This tail is more pronounced in low magnetic fields, indicating that these paramagnetic impurities can be saturated in moderate applied fields. To reduce the contributions from both paramagnetic and ferromagnetic impurities we extract the susceptibility $\chi_0$ instead from the isotherms of the magnetization $M(H)$.

\begin{figure}[ptb]
\centering
\includegraphics[width=\linewidth]{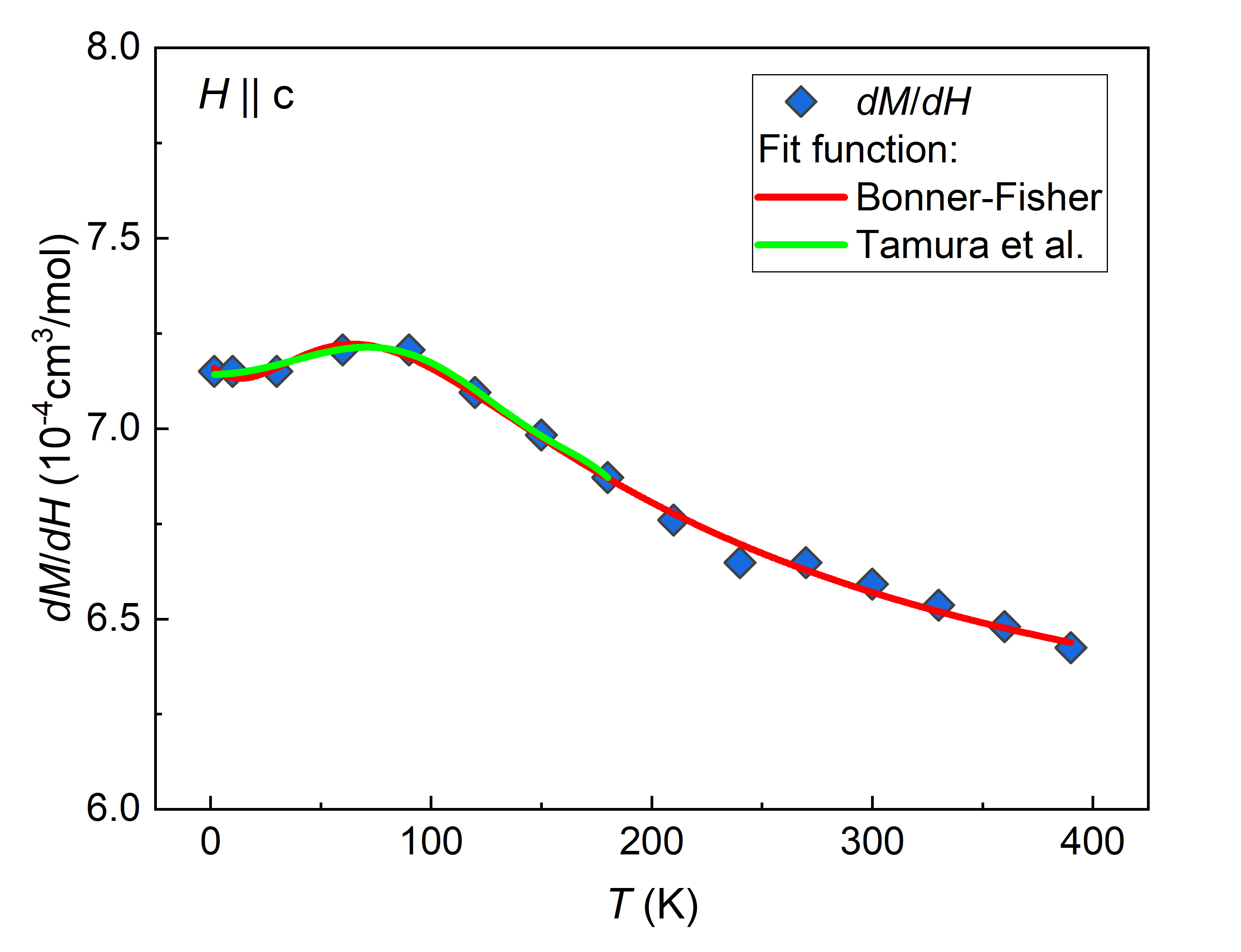}
\caption{Effective susceptibility $\chi_0 = dM/dH$ of La$_5$AgPb$_3$ as a function of temperature $T$, determined from $M(H)$ data as described in the main text. The magnetic field was applied parallel to the crystallographic c axis. Solid lines depict fits to the Bonner-Fisher model and the phenomenological expression from Ref.~\onlinecite{tamura1995}.}
\label{fig_MH}%
\end{figure}

$M(H)$ exhibits a linear dependence on magnetic field above \SI{10}{\kilo\\Oe}  at all measured temperatures [Fig.~\ref{fig_susT}(b)]. By extracting the slope of the M(H) curves from a linear fit to the data between \SIrange{10}{70}{\kilo\\Oe} we effectively remove all contributions by impurities that are fully magnetized in that field range, as they do not contribute to the slope of $M(H)$ anymore. The resulting susceptibility $\chi_0 = dM/dH$ is virtually identical to the corrected data depicted in Fig.~\ref{fig_susT}(main panel), but without the pronounced  impurity tail at low temperatures (Fig.~\ref{fig_MH}). As well, the observed weak maximum is shifted to slightly higher temperatures ($\sim$\SI{75}{\kelvin}). We note that this procedure does not remove contributions from impurities that are not fully magnetized in the selected fit range, which will be further discussed below. 

Intriguingly, similar maxima in the magnetic susceptibility have been observed in quasi-one dimensional conductors such as TTF-TCNQ (tetrathiafulvalene-tetracyanoquinodimethane)~\cite{torrance1977,scott1978}, and related systems~\cite{klotz1988,tamura1995}. The maxima are taken as evidence for electronic correlations, that as well result in an overall enhancement of the magnetic susceptibility beyond the value in a corresponding system with noninteracting electrons~\cite{torrance1977}. Tight-binding calculations and the Hubbard model adequately reproduce the high temperature part of the magnetic susceptibility in these quasi-one dimensional metals. In the limit of strong Coulomb interaction the magnetic part of the Hubbard model can in fact be equated to a Heisenberg linear chain of spins and it is found that the Bonner-Fisher (BF) model \cite{bonner1964linear} provides an adequate description of the measured susceptibility. 

We have, therefore, compared our corrected susceptibility $\chi_0$, to the following expression: $\chi_0 = \chi_\text{BF} + \chi_\text{CW} + \chi_\text{P}$, with $\chi_\text{BF}$ the polynomial expansion of the Bonner-Fisher model\cite{hatfield1981new}, $\chi_\text{CW} = C/(T-\theta)$ a Curie-Weiss term with Curie constant $C$ and temperature $\theta$ and $\chi_\text{P}$ the temperature independent Pauli susceptibility (see red line in Fig.~\ref{fig_MH}). Note, that we included a Curie-Weiss term to account for any remaining paramagnetic impurities that are not saturated in fields as large as 7 T.   The BF fit describes the data very well and leads to an effective moment of $\mu_\text{eff} = 0.50(1)\,\mu_\text{B}$ per formula unit, an exchange constant of $J/k_\text{b} = 61(4)$\,K and a Pauli susceptibility of $\chi_P = \SI{5.9\pm0.6}{\times10\tothe{-4}cm\tothe{3}\per\mol}$. The latter corresponds to a DOS $\mathcal{D}(E_\text{F}) = \SI{18.25\pm0.04}{states\per\electronvolt\per \text{f.u.}}$, which is in excellent agreement with the value determined from our specific heat data ($\SI{18.27\pm0.03}{states\per\electronvolt\per \text{f.u.}}$). Accordingly, the Wilson ratio $R_\text{W}= \chi_P/\gamma =1$ indicates that there is no ferromagnetic enhancement of the density of states in La$_{5}$AgPb$_{3}$. The overall magnitude of $\chi_0$ is enhanced by a factor of $\sim$ 2 compared to the expected Pauli susceptibility $\chi_P = \SI{3.1}{\times10\tothe{-4}cm\tothe{3}\per\mol}$, given our calculated density of states $\mathcal{D}(E_\text{F})=$ 9.5 states/eV/f.u. 

However, since we have no independent evidence that the Coulomb interaction has led to localized and moment-bearing electron states in La$_{5}$AgPb$_{3}$, we present as well a more phenomenological description~\cite{tamura1995}, where electronic correlations lead to spin fluctuations with a characteristic temperature scale T$^\ast$ that corresponds to the temperature where $\chi$ has a maximum, i.e. 75\,K in La$_{5}$AgPb$_{3}$. The fit to this model (Fig.~\ref{fig_MH}, green line) describes the data at low temperatures as well as the Bonner-Fisher expression. The observation of Fermi liquid behaviors at low temperature in the low temperature resistivity and specific heat, as well as the modest enhancements of the T$^{2}$ coefficient of the resistivity, the Sommerfeld coefficient, and the Pauli susceptibility all lend credence to this second interpretation. It is worth noting that the inferred spin fluctuations in La$_{5}$AgPb$_{3}$ have a significantly smaller scale T$^\ast$=75 K than those found in the quasi-one dimensional metals, where higher values of T$^\ast$ indicate significantly stronger electronic correlations.

 Overall, our analysis of the temperature dependence of the magnetic susceptibility lends further evidence that there are electronic states in La$_{5}$AgPb$_{3}$ with one-dimensional character, and as well moderate electronic correlations compared to other quasi-one dimensional systems that have been previously studied. We caution that not all of the states at the Fermi level are necessarily one-dimensional, but may instead contribute to the temperature independent Pauli susceptibility. 


\section{Conclusion}

We have, for the first time, synthesized high-quality single crystals of La$_{5}$AgPb$_{3}$ from flux and reported their physical properties, including magnetization, specific heat and electrical resistivity. Given the quasi-one dimensional character of the underlying crystal lattice, our objective was to assess La$_{5}$AgPb$_{3}$'s suitability as a benchmark system, being quasi-one dimensional, but without the substantial correlations that would generate localized magnetic moments or render it insulating. DFT computations revealed the presence of hybridized Ag atoms as well as electronically well isolated La$_2$ atoms that extend along the c-axis. Given the compositional flexibility of this structure type, these results are very encouraging that correlations and magnetism could be added, either by substituting moment-bearing transition metal atoms for Ag, or rare-earths for the La$_2$ atoms. 

The electronic structure calculations confirm expectations that La$_{5}$AgPb$_{3}$ has a robust and apparently conventional Fermi surface, and this result is verified by measurements of the electrical resistivity and the specific heat. Both indicate that La$_{5}$AgPb$_{3}$ displays Fermi Liquid behavior at the lowest temperatures. While this would not be expected for a truly one-dimensional system, it is rare that transport and thermal measurements find evidence for the Luttinger Liquid behavior that is expected~\cite{giamarchi2019clean}. The complexity of the Fermi surface, which is comprised of five different bands, raises a legitimate concern that not all of those bands are necessarily quasi-one dimensional. Spectroscopic investigations are required to definitively determine whether spinons and holons, the signature excitations of one-dimensional systems, are present in La$_{5}$AgPb$_{3}$.

Comparing the Sommerfeld coefficient of the specific heat to the values determined by the DFT computations indicates that the electronic correlations are moderate. In agreement, the temperature independent part of the magnetic susceptibility shows no enhancement relative to the density of states construed from the Sommerfeld constant, giving the Wilson constant R$_{W}$=1. This indicates that La$_{5}$AgPb$_{3}$ does not have significant ferromagnetic fluctuations. These results suggest that La$_{5}$AgPb$_{3}$ is a relatively uncorrelated metal, with a crystal structure that is suggestive of quasi-one dimensional character. The spectrum of one-dimensional systems that are currently known starts with the insulating spin chains, where the correlations are so strong that they localize electrons and form robust magnetic moments. Neutron scattering experiments find fractionalized excitations, proof of the electronic one-dimensionality of these systems. Fractionalized excitations are similarly observed in Yb$_{2}$Pt$_{2}$Pb~\cite{wu2016,gannon2019}, where the localized moments of the f-electrons of the Yb atoms form chains that are only weakly hybridized with the conduction electrons of this excellent metal. Decreasing the correlations leads to the delocalization of electrons and the collapse of magnetic moments at a Mott transition or crossover. The molecular conductors like TTF-TCNQ are apparently on the delocalized side of a Mott transition, where the extended states are still highly correlated. Provided that spectroscopic measurements find that La$_{5}$AgPb$_{3}$ is truly one-dimensional, it would be by far the most weakly correlated system yet identified. The flexibility of the $R_5MX_3$ ($R$ = rare earth, $M$ = transition metal or main group element, $X$ = Pb, Sn, Sb, In, Bi) family of compounds is quite promising that new compounds with intermediate degrees of correlations and magnetism can be explored.

\section{Acknowledgments}
We acknowledge support by the Natural Sciences and Engineering Research Council of Canada (NSERC). This research was supported by NSF-DMR-1807451. JM was supported in part by funding from the Max Planck-UBC-UTokyo Centre for Quantum Materials and the Canada First Research Excellence Fund, Quantum Materials and Future Technologies Program. We thank Ilya Elfimov for useful discussions.

\bibliography{references}



\end{document}